\documentclass[10pt,twocolumn,letterpaper]{article}

\usepackage{cvpr}
\usepackage{times}
\usepackage{graphicx}
\usepackage{booktabs}
\usepackage{amsmath}
\usepackage{amssymb}
\usepackage{latexsym}
\usepackage{soul}
\usepackage{csquotes}
\usepackage{xcolor}
\usepackage{url}

\usepackage[pagebackref=true,breaklinks=true,letterpaper=true,colorlinks,bookmarks=false]{hyperref}

\cvprfinalcopy 

\def\cvprPaperID{4} 

\ifcvprfinal\fi
\begin{document}

\title{Mosaic Super-resolution via Sequential Feature Pyramid Networks}
\author{
Mehrdad Shoeiby$^1$, Mohammad Ali Armin$^1$, Sadegh Aliakbarian$^{1,2}$, Saeed Anwar$^1$, Lars Petersson$^1$\\
$^1$CSIRO-Data61, $^2$Australian National University\\
{\tt\small firstname.lastname@data61.edu.au, sadegh.aliakbarian@anu.edu.au}
}

\maketitle

\begin{abstract}
Advances in the design of multi-spectral cameras have led to great interests in a wide range of applications, from astronomy to autonomous driving. However, such cameras inherently suffer from a trade-off between the spatial and spectral resolution. In this paper, we propose to address this limitation by introducing a novel method to carry out super-resolution on raw mosaic images, multi-spectral or RGB Bayer, captured by modern real-time single-shot mosaic sensors. To this end, we design a deep super-resolution architecture that benefits from a sequential feature pyramid along the depth of the network. This, in fact, is achieved by utilizing a convolutional LSTM (ConvLSTM) to learn the inter-dependencies between features at different receptive fields. Additionally, by investigating the effect of different attention mechanisms in our framework, we show that a ConvLSTM inspired module is able to provide superior attention in our context. Our extensive experiments and analyses evidence that our approach yields significant super-resolution quality, outperforming current state-of-the-art mosaic super-resolution methods on both Bayer and multi-spectral images. Additionally, to the best of our knowledge, our method is the first specialized method to super-resolve mosaic images, whether it be multi-spectral or Bayer.

\end{abstract}
 \vspace{-10pt}

\section{Introduction}
\label{sec:intro}


Real-time snap-shot mosaic image sensors are a category of imaging devices that encompass modern RGB and multi-spectral cameras. In fact, RGB cameras are a sub-category of multi-spectral cameras, only being capable of measuring three spectral channels \ie red, blue and green.  The recent improvements has given rise to multi-spectral cameras with the performance comparable to modern RGB cameras in terms of size, portability and speed~\cite{cdd_hyper_2013}.

Despite the great interest in these devices, with applications ranging from astronomy~\cite{msi_astronomy_2016} to tracking in autonomous vehicles \cite{takumi_2017,lars_2013}, they suffer from an inherent constraint: a trade-off between the spatial and the spectral \textit{resolution}. The reason is the limited physical space on 2D camera image sensors. A higher spatial resolution (smaller pixel size) reduces the number of possible wavelength channels on the image sensor, and thus creates a limitation in certain applications where the size and portability are essential factors, for instance, on a UAV~\cite{doering_2016}. A more portable (\ie smaller and lighter) camera suffers more from lower spatial and spectral resolution.  The spectral and spatial constraints of mosaic imaging devices motivates the need for super-resolution (SR) algorithms for the type of images that these devices create (mosaic images). Nevertheless, compared to the amount of existing literature on normal RGB (interpolated/demosaiced) RGB images SR, few efforts have been made toward mosaic image super-resolution (SR).

\begin{figure}
\begin{center}
\begin{tabular}{ccc}
\includegraphics[width=.14\textwidth]{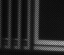}&
\includegraphics[width=.14\textwidth]{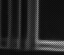}&
\includegraphics[width=.14\textwidth]{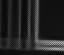}\\
a) & b) & c)\\ 
\end{tabular}
\end{center}
 \caption{Comparison between the baseline \cite{shi_2018} and our Mosaic super-resolution. a) Ground-truth, b) baseline, and c) our mosaic super-resolution. The baseline smooths out the bayers pattern and fuses the information while our method produces the results similar to the ground-truth}
 \label{fig:sample_results}
 \vspace{-17pt}
\end{figure}

The related literature aiming at the task of SR in RGB domain (discussed in Section~\ref{section:relate_work}) is predominantly carried out on interpolated/demosaiced RGB images. However, with most modern RGB cameras, mosaic Bayer images can be obtained instead of demosaiced images. In this regard, various studies~\cite{zhou_2018, fu_2018} have pointed out and discussed that interpolation or demosaicing deteriorates SR performance due to \textit{(1)} removing high-frequency information as interpolation/demosaicing can be viewed as a form of low pass filtering \cite{jaiswal2016adaptive}, while SR aims at predicting such high-frequency information; and \textit{(2)} interpolation/demosaicing introduces artifacts~\cite{zhou_2018,jaiswal2016adaptive} which can be either viewed as a signal loss or noise in the input image. The SR literature on mosaic RGB images, despite its importance, is limited to a few recent works \cite{ivrl_prime_2018,shoeiby_pirm2018_method,zhang_2018}. We believe that in most modern SR applications such as microscopy and astronomy, in which having access to high-resolution images is vital, it is counter-intuitive to throw away high-frequency content mosaic images and only rely on the interpolated/demosaiced images.


Surprisingly, despite its significance, mosaic image SR has been less considered in the literature. This, however, motivates us to conduct an in-depth study on how to get benefit of such vital information. Therefore, in this paper, we propose a SR framework for real-time mosaic image sensors (cameras) to bridge the gap identified in the literature. We believe that our approach will benefit many applications, such as in astronomy \cite{li2018super} or microscopy \cite{nehme2018deep} in which high quality super-resolved images are essential. To summarize, our primary contributions are:

\begin{itemize}
    \item We demonstrate an effective use of the ConvLSTM module to learn the relationship between features from different stages of a CNN with sequentially increasing receptive fields, and achieve state-of-the-art in mosaic SR.
    \item To the best of our knowledge, our method is the first addressing SR of mosaic images directly. We also demonstrate the different nature of mosaic images compared to demosaiced images by showing that methods specifically designed for mosaic SR does not  necessarily perform well on demosaiced/interpolated RGB images. 
\end{itemize}{}



As a secondary contribution, we experiment with different attention mechanisms and assess their relative performance in our network. Furthermore, investigating the structure of an LSTM module, we discover that elements of it are designed to apply implicit attention. By incorporating our LSTM inspired attention mechanism in our network, we empirically show its superior performance compared to other attention mechanisms.

\section{Related Work}
 \label{section:relate_work}
 
The RGB super-resolution methods dominate the SR literature; therefore, we first review the literature that focus on RGB SR and then discuss the more related existing literature on mosaic SR. One of the first works in RGB CNN based SR (SRCNN) \cite{dong2016SRCNNPAMI}, although simply composed of three convolutional layers, significantly outperformed the conventional SR algorithms. Following the success of SRCNN~\cite{dong2016SRCNNPAMI}, many CNN based algorithms~\cite{lim2017EDSR,dong2016FSRCNN,zhang2018SRMDNF} were developed. For example, fast SRCNN (FSRCNN)~\cite{dong2016FSRCNN} encompassing eight convolutional layers, speeds up the SR process by using as input the original low-resolution patch instead of a bi-cubically upsampled one. They highlight the fact that using interpolation to scale up images deteriorates the SR performance. Note that, extending from the discussion in \cite{dong2016FSRCNN}, RGB images are, in fact, interpolated from mosaic Bayer images, and hence, super-resolving directly from the raw mosaic images instead should result in better performance. 


Current approaches, similar to ours, use residual connections~\cite{kim2016VDSR,lim2017EDSR,ahn2018CARN}. For example, \cite{kim2016VDSR} introduced very deep SR (VDSR), which has a single global skip-connection from the input to the final output. Similarly, enhanced deep SR \ie EDSR~\cite{lim2017EDSR} employs residual blocks (RBs) with short skip connections. More recently, a cascading residual network (CARN)~\cite{ahn2018CARN} is proposed, which also employs a variant of RBs with cascading connections. While CARN~\cite{ahn2018CARN} is lagging behind EDSR~\cite{lim2017EDSR} in terms of PSNR, it improves efficiency and speed. More lately, motivated by the success of DenseNet~\cite{huang2017densely}, CNN-based SR networks have concentrated on the dense connection model. For example, SRDenseNet~\cite{tong2017image} uses dense connections to learn compact models, avoiding the problem of vanishing gradients and eases the flow from low-level features to high-level features.  Recently, the residual-dense network (RDN)~\cite{zhang2018RDN} employed dense connections to learn the local representations from the patches at hand. The dense network with multi-scale skip connections has similarities with our proposed method in terms of feature aggregation. However, their method aggregates features, whereas our method aggregates a sequence of features with sequentially increasing receptive fields, and uses a ConvLSTM module \cite{xingjian2015convolutional} to learn these sequential features.

The visual attention~\cite{mnih2014recurrent} concept in SR was introduced by RCAN~\cite{zhang_2018}, which models the inter-channel dependencies using a channel attention (CA) mechanism. This process is coupled with a very deep network composed of groups of RBs called $RG$s (RGs). Following in the footsteps of~\cite{zhang_2018}, the residual attention module (SRRAM) by~\cite{kim2018ram}, employed spatial attention as well as CA while still lagging behind RCAN~\cite{zhang_2018}. Most recently, Second-Order Attention Network (SAN) \cite{dai2019second} was introduced. The authors argue that the task of SR has achieved outstanding performance; however, at the expense of using very deep and wide networks. They argue that maybe a better utilization of intermediate features would help improve the results. Note that better utilization of intermediate features was brought up by RCAN; in fact, this was precisely the incentive behind CA in the RCAN setup.  Nevertheless, the SAN authors \cite{dai2019second} propose a second-order attention network within a residual in residual structure. The quantitative results are, on average, on par with RCAN, and in terms of the network size, SAN is only marginally smaller than RCAN (15.7M vs. 16M parameters).

The SR algorithms above focus mainly on super-resolving RGB images even though, as discussed before, the multi-spectral images are comparatively more adversely affected by the resolution constraints. Reviewing the limited multi-spectral/mosaic SR literature, one would realize that these networks are not structurally different \ie they are not fine-tuned for mosaic images by taking into account any spectral correlation of different channels. The scarcity of multi-spectral SR algorithms may be due to the absence of multi-spectral SR benchmarking platforms, as well as the difficulty of accessing suitable SR spectral datasets. For instance, \cite{li_sr_spectral_2017} aims to improve the quality of hyperspectral (not multi-spectral) images and is one of the few CNN based spectral SR methods.  To the best of our knowledge, the only multi-spectral SR methods~\cite{lahoud_2018,shi_2018} were submitted to the PIRM2018 multi-spectral SR challenge~\cite{shoeiby_pirm2018_method,shoeiby_dataset_2018}. The work in~\cite{lahoud_2018} adopted an image completion technique that requires $\times2$ and $\times3$ down-sampled images as input to a 12 layer convolutional network. While achieving good results, it addresses the problem of $\times3$ SR given $\times2$ and $\times3$ down-sampled images, rather than single-image SR. It is also not an end-to-end CNN based implementation. The best end-to-end CNN based method in the challenge was proposed by~\cite{shi_2018}, which implicitly employed the RCAN~\cite{zhang_2018}, which is the current state-of-the-art in multi-spectral SR.

The main body of the RCAN structure constitutes a sequence of RGs, with the receptive field increasing sequentially, that is, at a deeper RG it sees a larger area of the input image. This can be considered as different levels of an image pyramid. In one of our main contributions, we propose that a convolutional LSTM (ConvLSTM) \cite{xingjian2015convolutional} can learn the sequential relationship between these pyramidal feature levels. A superficially similar idea was used for optical flow estimation. To elaborate, the authors refer to \textit{pyramid convolutional LSTMs} \cite{guan2019unsupervised} in their structure, in which the pyramid part uses one convolutional LSTM at each semantic level to generate features by taking the number of input frames as the step size for the LSTMs. On the contrary, we only use one Convolutional LSTM with the step size being the number of semantic levels (RGs) that are being considered. Our second contribution is the use of a convolutional LSTM at the input of the network to learn the sequential relationship between different wavelengths of the network. Our third contribution is a self-attention mechanism that is inserted between the RGs of the RCAN structure.

\section{Method}
\label{sec:method}

\begin{figure*}
  \includegraphics[trim={0.5cm 1cm 1cm 1cm}, clip, width=\linewidth]{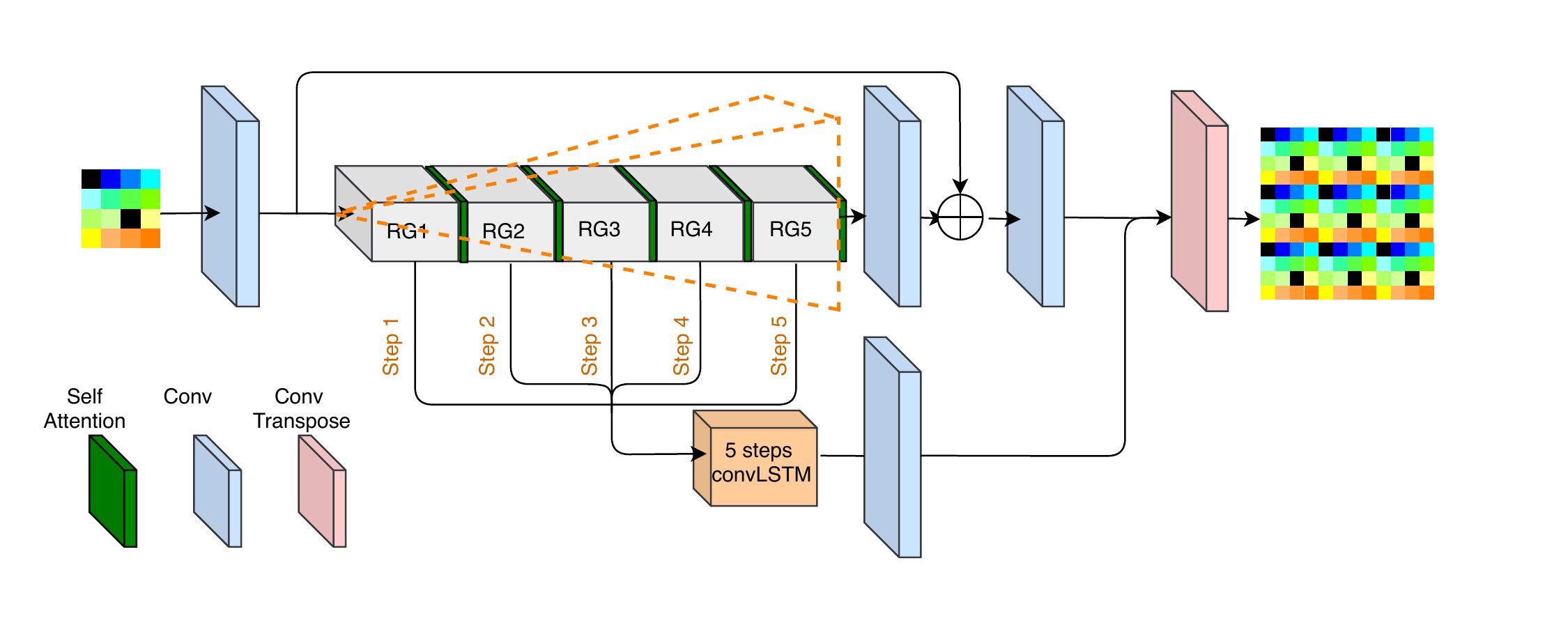}
  \caption{Overview of our proposed SR network.  Our model gets as input an $LR$ mosaic image and $\times 3 $super-resolves it to $HR$ mosaic image. Our network is based on RCAN \cite{shi_2018} with five residual groups ($RGs$) and three residual blocks ($RBs$).}
  \label{fig:network}
\end{figure*}


Inspired by the success of RCAN~\cite{shi_2018} in multi-spectral SR, we consider a simplified RCAN as the backbone for our method and develop our framework on top of it.
Briefly, the multi-spectral RCAN consists of five $RGs$ (RGs), and each RG has three $RBs$ and each $RB$ has one CA. We empirically observed that the features from each RG can be utilized better if higher level of aggregations are considered. In other word, we found that although processing features in a feed-forward manner has shown promising performance, one can better utilize the intermediate features if the dependencies between different RGs are taken into account. To this end, we treat the output of each RG as a separate representation, processing them in a pyramid Bidirectional ConvLSTM to learn relations between features of various receptive fields. We also observed that, despite the necessity for an attention mechanism, the CA used in the original RCAN cannot effectively learn to highlight the informative part of the feature vectors (as demonstrated in our experiments). Hence, we also design an attention mechanism, inspired by the internal operations of a ConvLSTM, and apply it between different $RG$s. In this section, we discuss different components of our model in detail.


 
\subsection{Bidirectional Pyramid ConvLSTM}
Bidirectional LSTMs have shown promising performance in learning long-range dependencies in a sequence. As the name implies, this class of models is capable of learning such relations in both the forward and backward directions, providing the model with stronger representation compared to its unidirectional counterpart. In our case, we propose to treat the output of each RG in our backbone as a sequence of feature maps. In other words, the features at different receptive fields act as the features at different time-steps. Since our feature maps carry strong spatial information, we utilize a bidirectional ConvLSTM \cite{xingjian2015convolutional}.  

A ConvLSTM takes a sequence (in our case pyramid receptive fields which are output features of the RCAN $RG$s $X_{t}$) as input and apply the following operation on them:
\begin{equation}
\begin{split}
    i_{t} &= \sigma (W_{xi} * X_{t} +W_{hi} * H_{t-1} + W_{ci} \odot C_{t-1} +b_{i})\\
    f_{t} &= \sigma (W_{xf} * X_{t} +W_{hf} * H_{t-1} + W_{cf} \odot C_{t-1} +b_{f})\\
    C_{t} &= f_{t} \odot C_{t-1} + i_{t} \odot tanh (W_{xc} * X_{t} + W_{hc} * H_{t-1} + b_{c})\\
    o_{t} &= \sigma (W_{xo} * X_{t} +W_{ho} * H_{t-1} + W_{co} \odot C_{t} +b_{o})\\
    H_{t} &= o_{t} \odot tanh(C_{t})
\end{split}
\end{equation}
where $i_{t}$, $f_{t}$ and $o_{t}$ are input, forget, and output gate of a ConvLSTM cell. $C_{t}$ denotes the cell state which is passed to the next ConvLSTM cell and $H_{t}$ indicates the output features of a ConvLSTM cell. Here  $*$, and $\odot$ refers to the convolution operation and Hadamard product. $\sigma$ is a Sigmoid function. 
Our Bidirectional ConvLSTM has five steps for the features of 5 $RGs$, and it maintains two sets of hidden and state cells per unit, for back and forth sequences. This allows Bidirectional ConvLSTM to have access to receptive field contexts in both directions and therefore increases the performance of the proposed network.

Since the features representing each time-step carry information at different receptive fields (with respect to the input), we consider our design of ConvLSTM as a \textit{Pyramid}, thus naming this component Pyramid ConvLSTM.

\begin{figure}
    \centering
    \includegraphics[width=0.5\textwidth]{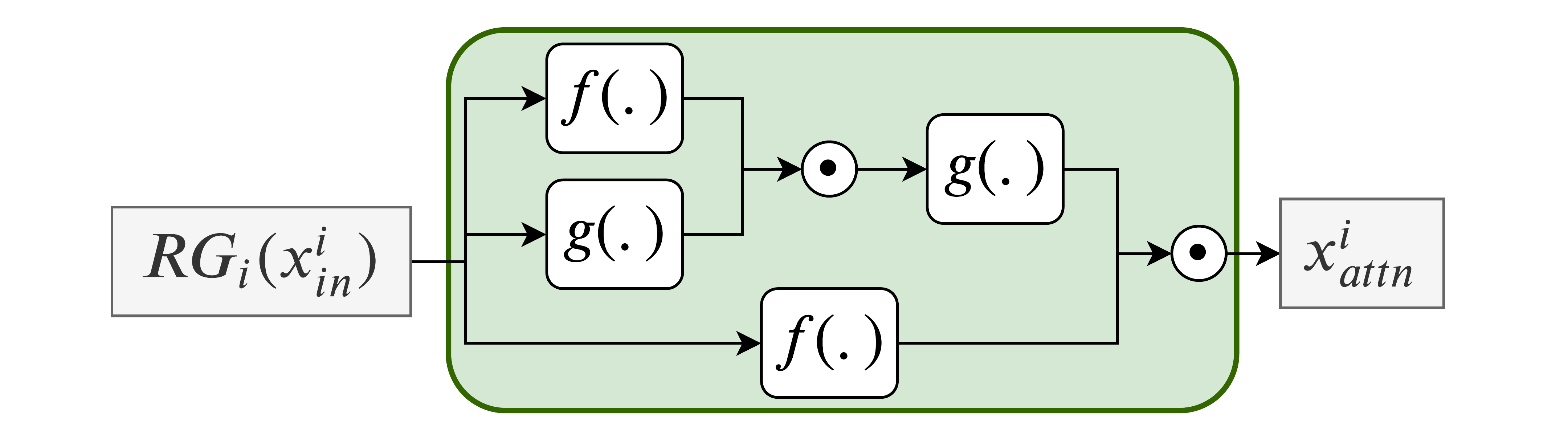}
    \caption{Overview of our self-attention mechanism: Adhering to LSTM intuition, $f(.)$ is a $Sigmoid$ function and $g(.)$ is a $Tanh$ function. }
    \label{fig:self_attn}
    \vspace*{-4mm}
\end{figure}{}
\vspace*{-3mm}

\subsection{Self attention mechanism}
As discussed earlier in Section~\ref{sec:method}, we expect an attention mechanism to play a tangible role in SR. Following this expectation, the original RCAN \cite{zhou_2018} architecture employs CAs inside $RG$s. When delving into the effect of each component in the backbone, we observed an inconsistency in the effect of CA in performance. Depending on the data presented, e.g., in mosaic format or cube format, Bayer or Multi-spectral, the effect of CA varies. Also, CA in RCAN is deployed 3 times in each $RG$. Therefore, an attention mechanism between the $RGs$ may be exploited as a way to achieve more efficient flow of information. This observation inspired us to investigate what type of attention mechanisms better suits the problem at hand. 

To this end, we first investigate existing attention mechanisms~\cite{zhang_2018, hu2018squeeze, fang2019bilinear} that have shown reasonable results on different computer vision problems. The failure of existing attention mechanisms in SR suggests that the nature of our problem is different from the ones cited above. By studying the LSTM structure \cite{hochreiter1997long} carefully, we realize that an LSTM by design provides implicit attention to input features and selectively passes more vital information to the next stage. The structure in Figure~\ref{fig:self_attn} is equivalent to an LSTM cell with only one step and with zero-initialized hidden and cell states. If we insert this structure between different $RGs$ (the green block in Figure~\ref{fig:network}) the first $Tanh$ followed by a $Sigmoid$ applies a non-linearity on the input feature map and then performs a gating operation, determining what information is needed to be passed to the next stage. We repeat this process twice to provide additional non-linearity and gating, which follows the intuition behind LSTM operations. To this end, and inspired by the gating inside a convolutional LSTM, our self-attention mechanism gets as input the output of each $RG$ and applies the following function:
\vspace*{-3mm}

\begin{equation}
\begin{split}
    x_{attn}^i  = &f(RG_i(x_{in}^i)) \odot \\
                  &g(f(RG_i(x_{in}^i) \odot g(RG_i(x_{in}^i))
\end{split}
\label{eq:attention}
\end{equation}
where $RG_i$ is the $i^{th}$ $RG$, $x_{in}^i$ is the input feature map to the $i^{th}$ $RG$, and $x_{attn}^i$ is the resulting feature map for its corresponding input. To stay with the logic behind LSTMs \cite{hochreiter1997long}, in our design, $f$ and $g$ are the non-linear functions of $Sigmoid$, and $Tanh$  respectively. 

As mentioned before, the attention mechanism shown in Eq.~\ref{eq:attention}, can be considered equivalent to the internal operations of a convolutional LSTM when the cell states carry no information. This is well-suited to our task since we do ignore any relation to other RGs and computing the refined features based only on the output of the current RG, acting as \textit{self}-attention. 

\subsection{Loss functions}

\vspace*{-3mm}

In SR literature, a simple loss function such as $L_1$~\cite{zhang_2018} or $L_2$~\cite{shi_2018,lahoud_2018}, or a perceptual loss function such as SSIM \cite{ssim_wang_2004} is usually utilized to train models. Here, for consistency, we choose $L_1$ loss as our baseline loss function since an $L_1$ function is less sensitive to outliers compared to an $L_2$ function. We use the PyTorch $SmoothL_1$  implementation, which is a more stable implementation compared to the vanilla $L_1$. $SmoothL_1$ can be expressed as  
\begin{equation}
SmoothL_{1} (\Theta) =\frac{1}{N}\sum_{i=1}^M Z^{i}
\label{eq:sid}
\end{equation}
where
 \[
    Z^{i}=
    \left\{
                \begin{array}{ll}
                  0.5\times (DIF)^2   &if |DIF|<1\\
                  |DIF|-0.5           &otherwise,
                \end{array}
           \right.
  \]
and $DIF =HR^{i}_{RGB}-LR^{i}_{MS}$. 

\section{Dataset generation}

\paragraph{Multi-spectral dataset.} We generate HR and LR mosaic images from HR multi-spectral cubes in the StereoMSI dataset~\cite{shoeiby_dataset_2018}, by taking the spatial offset of each wavelength in a multi-spectral pixel into account~\cite{shoeiby_dataset_2018}. The HR multi-spectral cubes have a dimension of $14 \times 240 \times 480$, and LR$\times3$ have a dimension of $14 \times 80 \times 160$. The multi-spectral images have 16 channels and each pixel exhibit a  $4\times4$ pattern \cite{shoeiby_dataset_2018}. However, as a result of the particular camera that captured these images, two of these channels are spectrally redundant and are not present, leaving us with 14 channels. Following the spatial location provided in \cite{shoeiby_dataset_2018}, we transform this 14 channel cube to a mosaic pattern. For the two redundant wavelengths, we assign zero value. In Figure~\ref{fig:network}, these two wavelengths are indicated by black pixels.  The resulting HR and LR mosaic patterns have dimensions $1 \times 960 \times 1920$, and $1 \times 320 \times 640$ respectively.

\vspace*{-4mm}

\paragraph{Bayer dataset} Regarding the Bayer dataset~\cite{shoeiby_dataset_2018}, an extended StereoMSI dataset has become available.  The size of the images is $1086\times 2046$.  To generate LR mosaic images, the HR mosaic was used to build an image cube of size $4\times 362 \times 682$. The $4$ channels correspond to two green, one red, and one blue. The image cubes were down-sampled and used to reconstruct LR mosaic images.
\vspace*{-4mm}
\paragraph{Converting multi-spectral mosaics to zero-padded multi-spectral cubes.} In a recent multi-spectral color-prediction work~\cite{shoeiby_color-predict_2019}, the authors proposed converting multi-spectral mosaics into the format of zero-padded multi-spectral cubes (for simplicity, we refer to this format as \textit{mosaic cubes}) as a way to give the network an extra wavelength dimension and they showed that this data representation helps achieve better performance; We verify that this data representation indeed helps us boost our quantitative results for multi-spectral SR. Please note that we use the cubic data format as the input, and the corresponding, actual, mosaic images are used as ground truth. In the case of Bayer SR, we did not observe any improvements. Although it remains to be demonstrated, the reason could be that Bayer pixels contain two green pixels with identical wavelengths. Hence, for Bayer SR we use simple mosaic images.
\section{Experiments}
\subsection{Settings}

\paragraph{Dataset:}We evaluate our approach on the PIRM2018 spectral SR challenge dataset~\cite{shoeiby_dataset_2018, shoeiby_pirm2018_method} as our multi-spectral mosaic SR evaluation. We use their extended Bayer images available for RGB mosaic SR evaluation. With the multi-spectral dataset, we have 350 $HR$-$LR$ image pairs with 300 images used for training and 30 and 20 images set aside for validation and testing, respectively. For the Bayer dataset, to stay within a comparable dataset size, we have $300$ training image pairs, $50$ for validation, and $50$ for testing.  

\vspace*{-4mm}
\paragraph{Evaluation metrics:}

 The 20 (multi-spectral) and 50 (Bayer) test images were super-resolved to a scale of $\times3$ and evaluated using the Pixel Signal to Noise Ratio (PSNR) and Structural Similarity Index (SSIM) metrics. For the SSIM metric, a window size of 7 is used with the metric being calculated for each channel and then averaged. 
\vspace*{-4mm}
\paragraph{Training settings:}
During training, we performed data augmentation on each mini-batch of 16 images which included random $60\times 60$ cropping of the input image, random rotation by $0^{\circ}$, $90^{\circ}$ ,$180^{\circ}$ , $270^{\circ}$ with $p=0.25$, and random horizontal flips with $p=0.5$. Our model is trained by the ADAM optimizer \cite{adam_2014} with $\beta_1 = 0.9$, $\beta_2=0.999$, and $\epsilon = 10^{-8}$. The initial learning rate was set to $10^{-4}$ and then halved every 2500 epochs. To implement our models, we used the PyTorch framework.
To test our algorithms, we select the models with the best performance on the validation set and present the test results for those models. 


\section{Results and Discussion}

\subsection{Multi-spectral mosaic SR}
\begin{table}
\centering
\caption{Mean and standard deviation (in parenthesis) of PSNR, SSIM on the  Multi-spectral dataset. Except where specified, the networks are input with zero-padded cubic mosaics. RCAN$^-$ indicates RCAN without CA.}
\tabcolsep=0.5cm
\scalebox{0.8}{
\begin{tabular}{lcc}
\toprule
\bf{Method}&\bf{PSNR} &\bf{SSIM} \\ \midrule
Bicubic             &   28.63     & 0.4552  \\ 
                    &   (3.50)    & (0.0691) \\  \midrule
RCAN  (Mosaic)      &   33.17    & 0.6500 \\ 
                    &   (3.62)    & (0.061) \\  \midrule
RCAN$^-$  (Mosaic)  &   33.15    & 0.6472 \\
                    &   (3.64)     & (0.0614) \\ \midrule
RCAN$^-$            &   33.16    & 0.6492 \\  
                    &   (3.65)     & (0.0616) \\ \midrule
RCAN                &   33.21    & 0.6500 \\
                    &   (3.64)     & (0.0610) \\ \midrule
PyrRCAN            &  \underline{33.293} & \underline{0.6535}\\ 
                    &  (3.70) & (0.0618)\\  \midrule
PyrRCAN + lstmA     & \bf{33.31}     & \bf{0.6522}  \\
                    & (0.0625)    &  (0.0625)\\  \bottomrule 
\end{tabular}}
\label{table:ms_results}
\end{table}
\begin{figure*}
  \includegraphics[trim={2cm 7cm 1.2cm 7cm},clip, width=\linewidth]{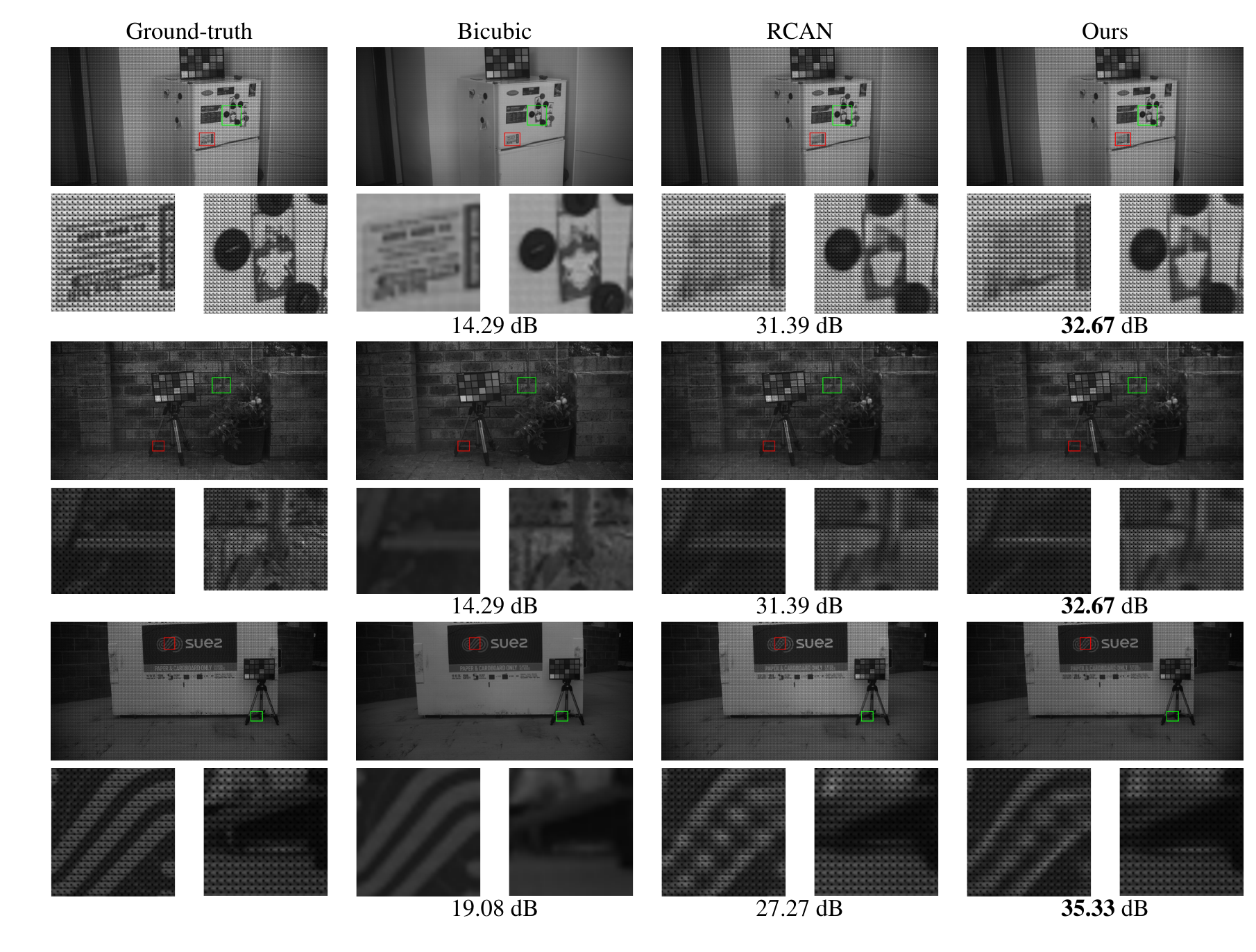}
    \vspace*{-8mm}
  \caption{Qualitative results for Bayer SR. Since the images are grayscale by nature, the results are best seen when zoomed in. Note, PSNR results per baseline are provided in the corresponding columns.}
  \label{fig:ms_fig}
  \vspace*{-4mm}
\end{figure*}
To assess the effect of using a mosaic vs. mosaic cube format~\cite{shoeiby_color-predict_2019}, In Table~\ref{table:ms_results}, we first train our baseline RCAN on mosaic data and mosaic cubes with $LR$ input (the mosaic format is always used for $HR$ ground truth), with and without CA (the first four rows). As explained before, and according to the results, the CA mechanism improves the performance more when using a mosaic cube data format compared with the mosaic data format. Moreover, the zero-padded cube data format improves the results compared to mosaic data by $0.04dB$. The rest of the experiments in Table~\ref{table:ms_results} are carried out with a zero-padded cube data format as the input, and mosaic data format as the output.  In all the tables, best and second-best results are shown in \textbf{bold} and \underline{underlined} fonts, respectively.

The fifth row of Table~\ref{table:ms_results} shows the effect of our Pyramid ConvLSTM structure, which has led to a considerable $0.08dB$ improvement in PSNR. The utilization of our proposed ConvLSTM inspired attention module, (lstmA), boosts the results by an additional $0.02dB$. In total, our proposed method boosts the SOTA approaches by $0.10dB$. Taking into account the effect of using mosaic cubes, our improvement adds up to $0.14dB$. Note that compared to the top PIRM2018 algorithms, our algorithm clearly outperforms existing methods. It is worth noting that no self-ensemble algorithm was used in the post-processing stage to achieve further improvements. These results purely demonstrate the effectiveness of our Pyramid ConvLSTM module boosted slightly by our lstmA module. Qualitative results, depicted in figure \ref{fig:ms_fig}, are also evident of the superiority of our method. 
\vspace*{-2mm}

\subsection{Bayer SR}
\begin{figure*}
  \includegraphics[trim={2cm 7cm 1.2cm 7cm},clip, width=\linewidth]{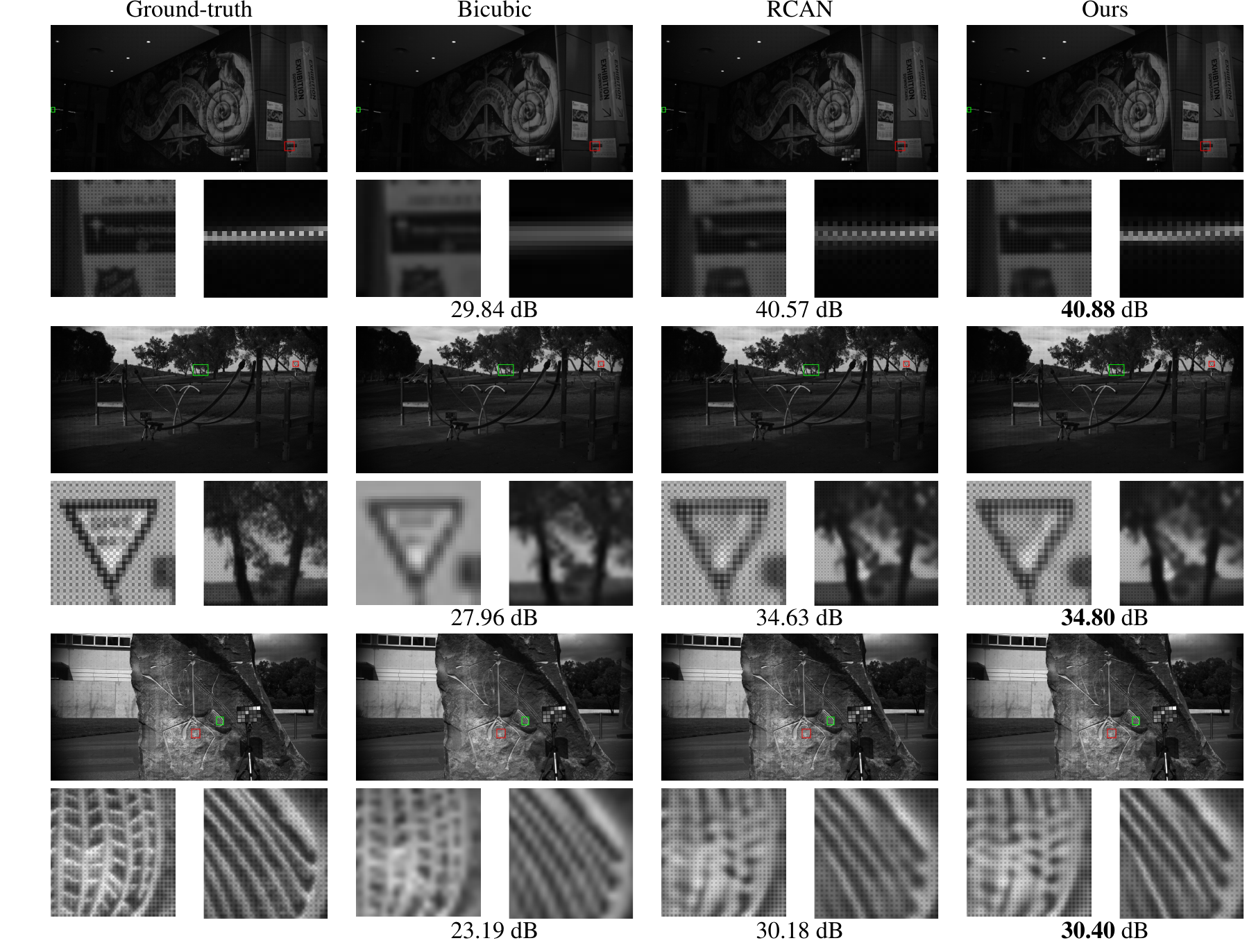}
  \caption{Qualitative results for Bayer SR. Since the images are grayscale by nature, the results are best seen when zoomed in. Note, PSNR results per baseline are provided in the corresponding columns.}
  \label{fig:bayer_fig}
  \vspace*{-4mm}
\end{figure*}

\vspace*{-2mm}
We use the mosaic data format for the Bayer SR task  based on our observation in which no improvement is obtained when using the mosaic cube data format. We hypothesize the reason is that Bayer pixels contain 2 green pixels with identical wavelengths, thereby defying the logic behind using mosaic cubes \cite{shoeiby_color-predict_2019}. The results are provided in Table~\ref{table:bayer_results}. The first two rows demonstrate the effect of CA, indicating that the model may not be able to take advantage of CA when uses Bayer data. Overall, our Pyramid ConvLSTM structure, together with the lstmA module, outperforms the baselines in terms of PSNR metric by $0.07dB$. Qualitative results, depicted in figure \ref{fig:bayer_fig}, are also evidence of the superiority of our method. Note, to the best of our knowledge, there are no specialized SR algorithms available on Bayer SR. Hence, we only compare with a bicubic baseline, which is customary in SR literature as well as the RCAN implementation \cite{shi_2018} that is SOTA in multi-spectral SR (RCAN \cite{zhang2018RCAN} is also SOTA in standard RGB SR). The closest algorithm to ours, as mentioned in section \ref{section:relate_work}, is \cite{zhou_2018}, which carries out joint demosaicing, and SR \ie does not produce mosaic images).
\begin{table}
\centering
\caption{Mean and standard deviation (in parenthesis) of PSNR, SSIM for the Bayer dataset. RCAN$^-$ indicates RCAN without CA.}
\tabcolsep=0.5cm
\scalebox{0.9}{
\begin{tabular}{lcc}
\toprule
\bf{Method}         &\bf{PSNR}        &\bf{SSIM} \\ \midrule
Bicubic             &   28.63        & 0.6398  \\ 
                    &   (3.50)       & (0.1364) \\  \midrule
RCAN                &   30.63        & 0.6560 \\ 
                    &   (3.65)       & (0.0998) \\ \midrule
RCAN$^-$            &   30.66        & 0.6589 \\
                    &   (3.65)       & (0.0988) \\\midrule       
PyrRCAN$^-$ + lstmA & \bf{30.70}           & \bf{0.6609}  \\ 
                    & (3.65)          & (0.1000)  \\ \bottomrule 
\end{tabular}}
\label{table:bayer_results}
\end{table}
 
\vspace{-10pt}

\subsubsection{Effect of ConvLSTM in PyrRCAN}
\begin{table}
\centering
\tabcolsep=0.5cm
\caption{Effect of Pyramid ConvLSTM}
\scalebox{0.9}{
\begin{tabular}{lcc}
\toprule
\bf{Method}&\bf{PSNR}   &\bf{SSIM} \\ 
\bottomrule
\multicolumn{3}{c}{Multi-spectral mosaic SR}\\
\toprule
PyrRCAN w/o ConvLSTM    &  33.21   & 0.6514\\ 
PyrRCAN                 &   \bf{33.29}  & \bf{0.6535} \\   
\bottomrule
\multicolumn{3}{c}{Bayer mosaic SR}\\
\toprule
PyrRCAN w/o ConvLSTM   &   30.591   & 0.655  \\ 
PyrRCAN                &  \bf{30.653}     & \bf{0.6582}\\
\bottomrule
\end{tabular}}
\label{table:convlstm_results}
\vspace{-10pt}
\end{table}

Here, we aim to assess whether the ConvLSTM is learning the sequence of the concatenated feature pyramid from features from different field of view, or it is merely the effect of reusing the intermediate features. We choose the PyrRCAN$^{-}$ structure, without our lstmA module, to better isolate the effect of the ConvLSTM module. We remove the ConvLSTM module and instead feed the features into the convolutional layer that follows the ConvLSTM in Fig.~\ref{fig:network}. The results, presented in Table~\ref{table:convlstm_results}, show the ConvLSTM module is indeed learning additional information from the concatenated features. In fact, without the ConvLSTM module, the results are worse than the baselines.

\subsubsection{Effect of attention}
\begin{table}
\centering
\caption{Ablation on different attention methods.}
\scalebox{0.8}{
\begin{tabular}{lcc}
\toprule
\bf{Method}&\bf{PSNR}   &\bf{SSIM} \\ 
\toprule
\multicolumn{3}{c}{Multi-spectral mosaic SR}\\
\bottomrule
PyrRCAN$^-$ + CA(RCAN)                             &  33.22   & 0.6512\\  
                                                   &  (3.6857)   & (0.06185)\\ 

PyrRCAN$^-$ + CA \cite{hu2018squeeze}            &   33.22   & 0.6511  \\ 
                                                 &   (3.67)   & (0.0622)  \\ 
PyrRCAN$^-$ + Bi-linear attention \cite{fang2019bilinear}  &   \underline{33.24}   & \bf{0.6517} \\ 
                                                        &   (3.67)   & (0.06277) \\ 
PyrRCAN$^-$ + lstm w/o Sigmoid                        &   33.22   & 0.6475 \\ 
                                                      &   (3.712)& (0.6475) \\
PyrRCAN$^-$ + lstmA                             & \bf{33.26}         & \underline{0.6513}  \\ 
                                                 & (3.70)         & (0.0622)  \\     \bottomrule
\multicolumn{3}{c}{Bayer mosaic SR}\\
\toprule
PyrRCAN$^-$ + CA(RCAN)                    &  30.65 & 0.6580\\  
                                           &  (3.6582) & (0.1001)\\ 
PyrRCAN$^-$ + CA \cite{hu2018squeeze}               &   30.63   & 0.6576  \\ 
                                           &   (3.69573)   & (0.09949) \\
PyrRCAN$^-$ + Bi-linear attention \cite{fang2019bilinear}   &   30.63   & 0.6546  \\ 
                                            &   (3.652)   & (0.0998) \\ 
PyrRCAN$^-$ + lstmA w/o Sigmoid             &   \underline{30.67}   & \underline{0.6612}  \\ 
                                            &   (3.664)   & (0.0998) \\ 
PyrRCAN$^-$ + lstmA                         & \bf{30.700}   & \bf{0.6609}  \\ 
                                         & (3.658)  & (0.1000)  \\ \bottomrule
\end{tabular}}
\label{table:atten_results}
\vspace{-10pt}
\end{table}
As discussed in Section~\ref{sec:method}, we investigate the effect of different attention mechanisms when placed between the $RGs$, guided by the intuition that such a mechanism can facilitate more effective information flow between the $RGs$. The attention mechanisms of our choice are \textit{(i)} the CA that was used in the SOTA multi-spectral SR work~\cite{shi_2018}, \textit{(ii)} the CA mechanism proposed in~\cite{hu2018squeeze}, \textit{(iii)} a bi-linear attention mechanism proposed in \cite{fang2019bilinear}, \textit{(iv)} our proposed lstmA without the $Sigmoid$ function in the bottom branch (to emulate a mask attention mechanism), and finally \textit{(v)} our proposed lstmA. To assess the effect of these mechanisms more directly, we remove CA from RCAN, so the networks under study only use one type of attention mechanism and only between $RGs$. The results show that our proposed lstmA mechanism outperforms all the other methods, and it is even marginally superior to BLA proposed in~\cite{fang2019bilinear}. The results of the ablation study for Bayer mosaic SR follow a more or less similar trend as the results on multi-spectral mosaic SR. 

\subsection{Demosaiced RGB images}
Our experimental observations indicated that our method is not as effective on demosaiced RGB images, \ie, standard RGB images, as it proved to be for mosaic images. The reason for this was discussed in Section~\ref{sec:method}. Demosaiced/interpolated images can, in fact, be considered low pass filtered, lacking some crucial information that can be exploited by an SR algorithm. We believe our Pyramid ConvLSTM structure is capable of exploiting such high-frequency information that may be discarded in the process of interpolation. For instance, sub-pixels (wavelengths) in mosaic Bayer and multi-spectral pixels display a high-frequency change in intensity, crucial information which is somewhat absent from an interpolated image.  Also, high-frequency patterns in either $2\times2$ or $4\time4$  pixels, which seems to appear throughout a mosaic image, contain some intra-wavelength dependencies typical to a multi-spectral or hyper-spectral pixel~\cite{msi_remote_2009}.  A sequential feature pyramid, as we have proposed, is capable of capturing these dependencies throughout a mosaic image.

\subsection{Comparison with other methods:}

Table \ref{table:other_methods} presents results using other modern CNN based SR methods for comparison. The methods we choose are VDSR \cite{shi_vd_sr_cnn_2016}, EDSR \cite{lim2017EDSR}, and CARN \cite{ahn2018CARN}. We train these networks, from the scratch, using our mosaic multi-spectral and RGB datasets.


\begin{table}
\centering
\caption{Comparison with different Multi-spectral mosaic and Bayer mosaic SR methods (in PSNR).}
\vspace{-10pt}
\scalebox{0.8}{
\begin{tabular}{lcccc}\\\hline
&\multicolumn{4}{c}{SR Methods}\\
\toprule
\bf{Datasets}& VDSR   &EDSR & CARN &  Ours  \\ 
\toprule
Multi-spectral & 30.61   & 31.06 &30.61&  \bf{33.26}  \\  \bottomrule
Bayer          & 29.68   & 30.03 &29.87& \bf{30.70}  \\  \bottomrule
\end{tabular}}
\label{table:other_methods}
\vspace{-10pt}
\end{table}
\section{Conclusion}

In this paper, we presented an SR algorithm designed explicitly for mosaic super-resolution. Our algorithm exhibit SOTA performance, achieved primarily via constructing a sequential feature pyramid and exploiting a ConvLSTM module to learn the inter-dependencies in the sequence. We also explored different attention modules, replacing CA in RCAN, and observed that an LSTM inspired attention mechanism provides the most significant improvement.

Apart from achieving SOTA and providing structural novelties introduced in this paper, we believe the most important message to convey is in regards to the task of Bayer SR. An intuitive observation, verified by our experiments, shows that indeed, mosaic and demosaiced images are different, and algorithms specific to each format need to be developed. Also, if a real-life application requires an SR algorithm, it makes more sense to capture Bayer images, which contain more high-frequency information, given that most modern RGB cameras are capable of outputting Bayer patterns. Hence, it is more beneficial to the computer vision community (\eg, microscopy, astronomy, food monitoring), that more research is dedicated to the task of mosaic super-resolution rather than standard RGB SR. We hope that this work encourages such a research~direction.

{\small
\bibliographystyle{ieee_fullname}
\bibliography{ref}

\begin{thebibliography}{10}\itemsep=-1pt

\bibitem{ahn2018CARN}
Namhyuk Ahn, Byungkon Kang, and Kyung-Ah Sohn.
\newblock Fast, accurate, and, lightweight super-resolution with cascading
  residual network.
\newblock {\em arXiv preprint arXiv:1803.08664}, 2018.

\bibitem{msi_astronomy_2016}
James~F Bell, Danika Wellington, Craig Hardgrove, Austin Godber, Melissa~S
  Rice, Jeffrey~R Johnson, and Abigail Fraeman.
\newblock Multispectral imaging of mars from the mars science laboratory
  mastcam instruments: Spectral properties and mineralogic implications along
  the gale crater traverse.
\newblock In {\em AAS/Division for Planetary Sciences Meeting Abstracts},
  volume~48, 2016.

\bibitem{dai2019second}
Tao Dai, Jianrui Cai, Yongbing Zhang, Shu-Tao Xia, and Lei Zhang.
\newblock Second-order attention network for single image super-resolution.
\newblock In {\em Proceedings of the IEEE Conference on Computer Vision and
  Pattern Recognition}, pages 11065--11074, 2019.

\bibitem{doering_2016}
D Doering, MR Vizzotto, C Bredemeier, CM da Costa, RVB Henriques, E Pignaton,
  and CE Pereira.
\newblock Mde-based development of a multispectral camera for precision
  agriculture.
\newblock {\em IFAC-PapersOnLine}, 49(30):24--29, 2016.

\bibitem{dong2016SRCNNPAMI}
Chao Dong, Chen~Change Loy, Kaiming He, and Xiaoou Tang.
\newblock Image super-resolution using deep convolutional networks.
\newblock {\em TPAMI}, 2016.

\bibitem{dong2016FSRCNN}
Chao Dong, Chen~Change Loy, and Xiaoou Tang.
\newblock Accelerating the super-resolution convolutional neural network.
\newblock In {\em ECCV}, 2016.

\bibitem{fang2019bilinear}
Pengfei Fang, Jieming Zhou, Soumava~Kumar Roy, Lars Petersson, and Mehrtash
  Harandi.
\newblock Bilinear attention networks for person retrieval.
\newblock In {\em Proceedings of the IEEE International Conference on Computer
  Vision}, pages 8030--8039, 2019.

\bibitem{fu_2018}
Ying Fu, Yinqiang Zheng, Hua Huang, Imari Sato, and Yoichi Sato.
\newblock Hyperspectral image super-resolution with a mosaic rgb image.
\newblock {\em IEEE Transactions on Image Processing}, 27(11):5539--5552, 2018.

\bibitem{msi_remote_2009}
Alexander~FH Goetz.
\newblock Three decades of hyperspectral remote sensing of the earth: A
  personal view.
\newblock {\em Remote Sensing of Environment}, 113:S5--S16, 2009.

\bibitem{guan2019unsupervised}
Shuosen Guan, Haoxin Li, and Wei-Shi Zheng.
\newblock Unsupervised learning for optical flow estimation using pyramid
  convolution lstm.
\newblock In {\em 2019 IEEE International Conference on Multimedia and Expo
  (ICME)}, pages 181--186. IEEE, 2019.

\bibitem{hochreiter1997long}
Sepp Hochreiter and J{\"u}rgen Schmidhuber.
\newblock Long short-term memory.
\newblock {\em Neural computation}, 9(8):1735--1780, 1997.

\bibitem{hu2018squeeze}
Jie Hu, Li Shen, and Gang Sun.
\newblock Squeeze-and-excitation networks.
\newblock In {\em Proceedings of the IEEE conference on computer vision and
  pattern recognition}, pages 7132--7141, 2018.

\bibitem{huang2017densely}
Gao Huang, Zhuang Liu, Laurens van~der Maaten, and Kilian~Q Weinberger.
\newblock Densely connected convolutional networks.
\newblock In {\em CVPR}, 2017.

\bibitem{jaiswal2016adaptive}
Sunil~Prasad Jaiswal, Lu Fang, Vinit Jakhetiya, Jiahao Pang, Klaus Mueller, and
  Oscar~Chi Au.
\newblock Adaptive multispectral demosaicking based on frequency-domain
  analysis of spectral correlation.
\newblock {\em IEEE Transactions on Image Processing}, 26(2):953--968, 2016.

\bibitem{kim2016VDSR}
Jiwon Kim, Jung Kwon~Lee, and Kyoung Mu~Lee.
\newblock Accurate image super-resolution using very deep convolutional
  networks.
\newblock In {\em CVPR}, 2016.

\bibitem{kim2018ram}
Jun-Hyuk Kim, Jun-Ho Choi, Manri Cheon, and Jong-Seok Lee.
\newblock Ram: Residual attention module for single image super-resolution.
\newblock {\em arXiv preprint arXiv:1811.12043}, 2018.

\bibitem{adam_2014}
Diederik~P Kingma and Jimmy Ba.
\newblock Adam: A method for stochastic optimization.
\newblock {\em arXiv preprint arXiv:1412.6980}, 2014.

\bibitem{ivrl_prime_2018}
Fayez Lahoud, Ruofan Zhou, and Sabine S\"usstrunk.
\newblock Multi-modal spectral image super-resolution.
\newblock In {\em Proceedings of the European Conference on Computer Vision
  (ECCV)}, 2018.

\bibitem{lahoud_2018}
Fayez Lahoud, Ruofan Zhou, and Sabine S{\"u}sstrunk.
\newblock Multi-modal spectral image super-resolution.
\newblock In {\em European Conference on Computer Vision}, pages 35--50.
  Springer, 2018.

\bibitem{li_sr_spectral_2017}
Yunsong Li, Jing Hu, Xi Zhao, Weiying Xie, and JiaoJiao Li.
\newblock Hyperspectral image super-resolution using deep convolutional neural
  network.
\newblock {\em Neurocomputing}, 266:29--41, 2017.

\bibitem{li2018super}
Zhan Li, Qingyu Peng, Bir Bhanu, Qingfeng Zhang, and Haifeng He.
\newblock Super resolution for astronomical observations.
\newblock {\em Astrophysics and Space Science}, 363(5):92, 2018.

\bibitem{lim2017EDSR}
Bee Lim, Sanghyun Son, Heewon Kim, Seungjun Nah, and Kyoung~Mu Lee.
\newblock Enhanced deep residual networks for single image super-resolution.
\newblock In {\em CVPRW}, 2017.

\bibitem{mnih2014recurrent}
Volodymyr Mnih, Nicolas Heess, Alex Graves, et~al.
\newblock Recurrent models of visual attention.
\newblock In {\em Advances in neural information processing systems}, pages
  2204--2212, 2014.

\bibitem{lars_2013}
Mohammad Najafi, Sarah~Taghavi Namin, and Lars Petersson.
\newblock Classification of natural scene multi spectral images using a new
  enhanced crf.
\newblock In {\em 2013 IEEE/RSJ International Conference on Intelligent Robots
  and Systems}, pages 3704--3711. IEEE, 2013.

\bibitem{nehme2018deep}
Elias Nehme, Lucien~E Weiss, Tomer Michaeli, and Yoav Shechtman.
\newblock Deep-storm: super-resolution single-molecule microscopy by deep
  learning.
\newblock {\em Optica}, 5(4):458--464, 2018.

\bibitem{shi_vd_sr_cnn_2016}
Wenzhe Shi, Jose Caballero, Ferenc Husz{\'a}r, Johannes Totz, Andrew~P Aitken,
  Rob Bishop, Daniel Rueckert, and Zehan Wang.
\newblock Real-time single image and video super-resolution using an efficient
  sub-pixel convolutional neural network.
\newblock In {\em Proceedings of the IEEE Conference on Computer Vision and
  Pattern Recognition}, pages 1874--1883, 2016.

\bibitem{shi_2018}
Zhan Shi, Chang Chen, Zhiwei Xiong, Dong Liu, Zheng-Jun Zha, and Feng Wu.
\newblock Deep residual attention network for spectral image super-resolution.
\newblock In {\em Proceedings of the European Conference on Computer Vision
  (ECCV)}, pages 0--0, 2018.

\bibitem{shoeiby_color-predict_2019}
Mehrdad Shoeiby, Petersson Lars, Sadegh Aliakbarian, Ali Armin, and Antonio
  Robles-kelly.
\newblock Super-resolved chromatic mapping of snapshot mosaic image sensors via
  atexture sensitive residual network.

\bibitem{shoeiby_pirm2018_method}
Mehrdad Shoeiby, Antonio Robles-Kelly, Radu Timofte, Ruofan Zhou, Fayez Lahoud,
  Sabine Süsstrunk, Zhiwei Xiong, Zhan Shi, Chang Chen, Dong Liu, Zheng-Jun
  Zha, Feng Wu, Kaixuan Wei, Tao Zhang, Lizhi Wang, Ying Fu, Zhiwei Zhong,
  Koushik Nagasubramanian, Asheesh~K. Singh, Arti Singh, Soumik Sarkar, and
  Ganapathysubramanian Baskar.
\newblock {PIRM2018} challenge on spectral image super-resolution: Methods and
  results.
\newblock In {\em European Conference on Computer Vision Workshops (ECCVW)},
  2018.

\bibitem{shoeiby_dataset_2018}
Mehrdad Shoeiby, Antonio Robles-Kelly, Ran Wei, and Radu Timofte.
\newblock Pirm2018 challenge on spectral image super-resolution: Dataset and
  study.
\newblock {\em arXiv preprint arXiv:1904.00540}, 2019.

\bibitem{takumi_2017}
Karasawa Takumi, Kohei Watanabe, Qishen Ha, Antonio Tejero-De-Pablos, Yoshitaka
  Ushiku, and Tatsuya Harada.
\newblock Multispectral object detection for autonomous vehicles.
\newblock In {\em Proceedings of the on Thematic Workshops of ACM Multimedia
  2017}, pages 35--43. ACM, 2017.

\bibitem{tong2017image}
Tong Tong, Gen Li, Xiejie Liu, and Qinquan Gao.
\newblock Image super-resolution using dense skip connections.
\newblock In {\em ICCV}, 2017.

\bibitem{ssim_wang_2004}
Zhou Wang, Alan~C Bovik, Hamid~R Sheikh, and Eero~P Simoncelli.
\newblock Image quality assessment: from error visibility to structural
  similarity.
\newblock {\em IEEE transactions on image processing}, 13(4):600--612, 2004.

\bibitem{cdd_hyper_2013}
Di Wu and Da-Wen Sun.
\newblock Advanced applications of hyperspectral imaging technology for food
  quality and safety analysis and assessment: A review—part i: Fundamentals.
\newblock {\em Innovative Food Science \& Emerging Technologies}, 19:1--14,
  2013.

\bibitem{xingjian2015convolutional}
SHI Xingjian, Zhourong Chen, Hao Wang, Dit-Yan Yeung, Wai-Kin Wong, and
  Wang-chun Woo.
\newblock Convolutional lstm network: A machine learning approach for
  precipitation nowcasting.
\newblock In {\em Advances in neural information processing systems}, pages
  802--810, 2015.

\bibitem{zhang2018SRMDNF}
Kai Zhang, Wangmeng Zuo, and Lei Zhang.
\newblock Learning a single convolutional super-resolution network for multiple
  degradations.
\newblock In {\em CVPR}, 2018.

\bibitem{zhang_2018}
Yulun Zhang, Kunpeng Li, Kai Li, Lichen Wang, Bineng Zhong, and Yun Fu.
\newblock Image super-resolution using very deep residual channel attention
  networks.
\newblock In {\em European Conference on Computer Vision}, pages 294--310.
  Springer, 2018.

\bibitem{zhang2018RCAN}
Yulun Zhang, Kunpeng Li, Kai Li, Lichen Wang, Bineng Zhong, and Yun Fu.
\newblock Image super-resolution using very deep residual channel attention
  networks.
\newblock {\em arXiv preprint arXiv:1807.02758}, 2018.

\bibitem{zhang2018RDN}
Yulun Zhang, Yapeng Tian, Yu Kong, Bineng Zhong, and Yun Fu.
\newblock Residual dense network for image super-resolution.
\newblock In {\em CVPR}, 2018.

\bibitem{zhou_2018}
Ruofan Zhou, Radhakrishna Achanta, and Sabine S{\"u}sstrunk.
\newblock Deep residual network for joint demosaicing and super-resolution.
\newblock In {\em Color and Imaging Conference}, volume 2018, pages 75--80.
  Society for Imaging Science and Technology, 2018.

\end{thebibliography}
}

\end{document}